\documentclass[twocolumn,prl]{revtex4}
\usepackage{graphicx}
\usepackage{amsmath}
\usepackage{amssymb}
\usepackage{bm}

\begin{document}
%
\title{Berry phase, topology, and diabolicity in quantum nano-magnets}

\author{Patrick Bruno}\email{bruno@mpi-halle.de}
\affiliation{Max-Planck-Institut f{\"u}r Mikrostrukturphysik,
Weinberg 2, D-06120 Halle, Germany} \pacs{}

\date{\today}

\begin{abstract}
A topological theory of the diabolical points (degeneracies) of
quantum magnets is presented. Diabolical points are characterized
by their \emph{diabolicity index}, for which topological sum rules
are derived. The paradox of the the missing diabolical points for
Fe$_8$ molecular magnets is clarified. A new method is also
developed to provide a simple interpretation, in terms of
destructive interferences due to the Berry phase, of the
\emph{complete} set of diabolical points found in biaxial systems
such as Fe$_8$.
\end{abstract}

\maketitle
%
The energy levels of a quantum mechanical system generally tend to
repel each other, so that degeneracies constitute exceptional
events (in a sense to be specified below) \cite{vonNeumann1929}.
Such degeneracies, called diabolical points \cite{Berry1984b},
have recently attracted great attention in molecular magnets
\cite{Sessoli1993}, where they occur as a result of destructive
interference (due to the geometric Berry phase \cite{Berry1984a})
between different tunnelling paths \cite{Loss1992}, and give rise
to oscillations of the tunnel splitting of the ground state of
quantum magnets \cite{Garg1993}. Experiments on Fe$_8$ molecular
magnets have not only confirmed this \cite{Wernsdorfer1999}, but
also revealed the existence of further series of diabolical
points, which, so far, could not be understood in terms of
destructive interference due to the geometric phase. Furthermore,
some expected diabolical are missing, due to higher order
anisotropies \cite{Wernsdorfer1999}. In this paper, a general
topological theory of the diabolical points of quantum magnets is
presented. Diabolical points are characterized by their
\emph{diabolicity index}, for which topological sum rules are
derived. The paradox of the the missing diabolical points for
Fe$_8$ molecular magnets is clarified. A new method is also
developed to provide a simple interpretation, in terms of
destructive interferences due to the Berry phase, of the
\emph{complete} set of diabolical points found in biaxial systems
such as Fe$_8$.

The question of diabolical points, to be addressed in the present
paper, goes back to the famous von Neumann-Wigner theorem
\cite{vonNeumann1929} stating that, in a family of
parameter-dependent hermitian Hamiltonians, accidental
degeneracies of two successive eigenvalues are found on
submanifolds of codimension 3 of the parameter manifold. In other
words, if a hermitian Hamiltonian depends on 3 external real
parameters, such as the 3 components of the magnetic field,
degeneracies can be found only for isolated values of the magnetic
field, and therefore constitute a set of measure zero. Because the
double-cone shape of the eigenenergy surfaces near such
degeneracies resemble the toy called \emph{diabolo}, they have
been dubbed \emph{diabolical points} \cite{Berry1984b}.

\begin{figure}[b]
\begin{center}
\includegraphics[width=1.0\columnwidth]{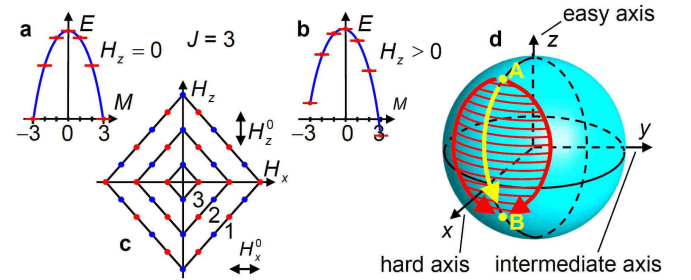}
\end{center}
\caption{\textbf{a},\textbf{b}: Schematic level diagram of a
biaxial spin system with $J\!=\!3$ and $0\!<\!D\!\ll\! K$ for
$H_z\!=\!0$ (\textbf{a}) and $H_z \!>\!0$ (\textbf{b});
\textbf{c}: diabolical points for a spin $J\!=\!3$ with biaxial
anisotropy; \textbf{d}: sketch of the various tunnelling paths
between states $A$ and $B$.} \label{fig_diabo}
\end{figure}

Interest in diabolical points has been renewed as Berry
\cite{Berry1984a} pointed out that they behave as magnetic
monopoles in parameter space, i.e., that a system that is
adiabatically transported around a closed circuit in parameter
space near a diabolical point acquires a phase shift (the Berry
phase) proportional to the solid angle of the circuit as seen from
the diabolical point. For quantum spin systems, it has been
pointed out \cite{Loss1992} that the occurrence of a diabolical
point implied by Kramers' theorem \cite{Kramers1930}, namely the
absence of tunnelling between degenerate ground states of
anisotropic quantum magnets of half-integer spin in zero field,
can be understood as due to destructive interference between
equivalent tunnelling paths whose Berry phase differ by an odd
multiple of $\pi$. Further, Garg \cite{Garg1993} has pointed out
that, as a magnetic field is applied along a hard axis, the solid
angle $\Omega$ enclosed between the two equivalent tunnelling
paths joining the classical ground states $A$ and $B$ (red arrows
in Fig.~\ref{fig_diabo}\textbf{d}) decreases from $2\pi$ to 0 with
increasing magnetic field, giving rise to $2J$ equidistant
diabolical point located on the hard axis (red dots at $H_z=0$ in
Fig.~\ref{fig_diabo}\textbf{c}). This prediction has been
confirmed in a beautiful experiment by Wernsdorfer and Sessoli
\cite{Wernsdorfer1999}, who observed this oscillatory behavior
(with 4 diabolical points on the positive hard axis) for the
spin-10 molecular magnet
$[(\mathrm{tacn})_6\mathrm{Fe}_8\mathrm{O}_2(\mathrm{OH})_{12}]^{8+}$
(usually abbreviated as Fe$_8$). Furthermore, Wernsdorfer and
Sessoli discovered, for non-zero values of the easy-axis field
$H_z$ (Fig.~\ref{fig_diabo}\textbf{b}), further series of
unexpected diabolical points, displaying a characteristic parity
alternation (red vs. blue points, on
Fig.~\ref{fig_diabo}\textbf{c}).

Following this discovery, the complete set of diabolical points
has been identified by semi-classical, perturbative, or algebraic
methods \cite{Garg1999,Villain2000,Kececioglu2001}. Writing the
biaxial Hamiltonian as $\mathcal{\hat{H}} \!=\!\mathcal{\hat{H}}_0
- \mathbf{H\cdot \hat{J}}$, with
$\mathcal{\hat{H}}_0 \! =\! -K \! \hat{J}_z^2 + D\!\left(
\hat{J}_x^2 \! - \hat{J}_y^2 \! \right)$,
and $0\!<\!D\!<\!K$, the diabolical points, corresponding to
degeneracies between the states $M$ and $-M'$ (labelling as in
Fig.~\ref{fig_diabo}\textbf{a},\textbf{b}), are exactly given by
\cite{Kececioglu2001}
\begin{subequations}
\begin{eqnarray}\label{eq_diabo_biaxial}
H_z&=&(M-M')H_z^0 , \\
H_x&=&\left(\frac{M+M'-1}{2}-n\right)H_x^0 ,
\end{eqnarray}
\end{subequations}
with
$n\!=\!0,1, \ldots, (M\!+\!M'\!-1)$,
$H_x^0\!\equiv\!2\sqrt{2D(K\!+\!D)}$
and
$H_z^0\!\equiv\! \sqrt{K^2\!-\!D^2}$.
The full set of diabolical points (for $J=3$) is shown in
Fig.~\ref{fig_diabo}\textbf{c}. One should note that several
diabolical points may coincide; the number of such coincident
diabolical points is the same on a given diamond, as indicated on
Fig.~\ref{fig_diabo}\textbf{c} \cite{Kececioglu2001}. In spite of
the striking apparent similarity between the sets of diabolical
points found on and off the hard axis, respectively, the latter
could not be interpreted in terms of destructive interferences
between tunnelling paths, which is very unsatisfactory.
Furthermore, only 4 diabolical points were observed on the
positive hard axis, instead of the 10 predicted, which has been
explained as due to a small tetragonal anisotropy term
\cite{Wernsdorfer1999}. However, what happens with the missing
diabolical points remains mysterious.

The general problem of diabolical points of quantum magnets may be
formulated as follows. We consider a spin-$J$ system with
Hamiltonian
$\mathcal{\hat{H}}\!=\!\mathcal{\hat{H}}_0(\mathbf{\hat{J}}) -
\mathbf{H\cdot \hat{J}}$,
where the zero-field Hamiltonian
$\mathcal{\hat{H}}_0(\mathbf{\hat{J}})$
is an arbitrary even function of the vector spin operator
$\mathbf{\hat{J}}$.
Note that the above Hamiltonian encompasses also the case of an
arbitrary tensorial $g$-factor, which can be accounted for by
properly rescaling the field components along the principal axes
of the $g$-tensor. To discuss the properties of diabolical points,
we take the convention to identify the $2J+1$ eigenstates by the
label $\mu$ running from $+J$ for the state of lowest energy to
$-J$ for the state of highest energy, with increments of 1 (this
labelling corresponds to the quantum number $M$ of $J_z$ for
$\mathbf{H}$ in the $+\mathbf{z}$ direction, and
$\mathcal{H}_0\rightarrow 0$). We call a diabolical point of order
$g$ (or a $g$-diabolical point), a point in $\mathbf{H}$-space
where $g$ successive eigenstates are degenerate; the various
diabolical points are labelled by an index $i$, running from $1$
to $N_d$, the total number of diabolical points corresponding to a
given Hamiltonian $\mathcal{\hat{H}}_0$. Note that several
diabolical points involving different sets of levels may coincide
in $\mathbf{H}$-space; this occurs, for instance, for the
diabolical points of the biaxial case mentioned above. We also
note, in passing, that the argument used by von Neumann and Wigner
to discuss the occurrence of 2-diabolical points can be
immediately generalized to show that the submanifolds on which we
find the coincidence of $n$ diabolical points of respective orders
$g_i$ $(1\leq i \leq n)$
are of codimension
$d\equiv \sum_{i=1}^n (g_i^2 -1)$.
A diabolical point where the levels $\mu$ to $\mu'$ (with
$\mu' \!< \mu$) are degenerate will be noted
$\mathbf{H}_{i\,(\mu)}^{(\mu')}$.

As discovered by Berry \cite{Berry1984a}, a quantum system in a
non-degenerate eigenstate $\mu$ adiabatically transported around a
closed curve $\mathcal{C}$ in parameter space (the external
magnetic field) acquires a geometric phase given by the flux
through a surface $\Sigma$ subtended by the circuit $\mathcal{C}$
of the Berry curvature
$\mathbf{B}_{(\mu)} \!\equiv\! -\mathrm{Im}
{\sum^\prime_{\mu^\prime}} \frac{\left\langle \mu \right|
\mathbf{\hat{J}} \left| \mu^\prime \right\rangle \times
\left\langle \mu^\prime \right| \mathbf{\hat{J}} \left| \mu
\right\rangle}{(E_\mu - E_{\mu^\prime})^2} ,$
where the sum is restricted to $\mu'\!\neq\!\mu$. The Berry
curvature $\mathbf{B}_{(\mu)}$ is divergenceless, except at
diabolical points involving the level $\mu$, where monopole
sources are located \cite{Berry1984a}. To each diabolical point,
we can associate a closed surface $\Sigma_i$ surrounding it, such
that (except for coinciding diabolical points) no other diabolical
point is enclosed inside $\Sigma_i$. For a diabolical point
$\mathbf{H}_{i\,(\mu_1)}^{(\mu_2)}$,
the definiteness of the wavefunction implies that the flux through
$\Sigma_i$ of $\mathbf{B}_{(\mu)}$
is topologically quantized, i.e., for
$\mu_2 \!\leq\! \mu\!\leq\! \mu_1$,
$Q_{i(\mu)}\!\equiv\! \frac{-1}{2\pi}\int_{\Sigma_i} \!
\mathbf{B}_{(\mu)}\!\! \cdot \!\mathrm{d}\mathbf{S} \in
\mathbb{Z};$
(by convention, we define
$Q_{i(\mu)}\!\equiv\! 0$ for $\mu\!
<\!\mu_2$ or $\mu\!>\!\mu_1$).
The \emph{topological charge} $Q_{i(\mu)}$ is known as a Chern
number, an analogous to the Euler index of a surface in
differential geometry. For a given diabolical point $i$, one can
prove the following sum rule (minor extension of a result of
\cite{Berry1984a}):
$\sum_\mu Q_{i(\mu)}\!=\!0 .$
Further, considering a surface $\Sigma$ enclosing all the
diabolical points (one can show easily that such a surface
exists), the flux of the Berry curvature through $\Sigma$, being a
topological invariant, should remained unchanged as
$\mathcal{\hat{H}}_0$ is scaled down to zero, and is therefore
given by the Chern number of a spin in a Zeeman field, which
yields another sum rule:
$\sum_i Q_{i(\mu)} \!= 2\mu .$

\begin{figure}[t]
\begin{center}
\includegraphics[width=0.8\columnwidth]{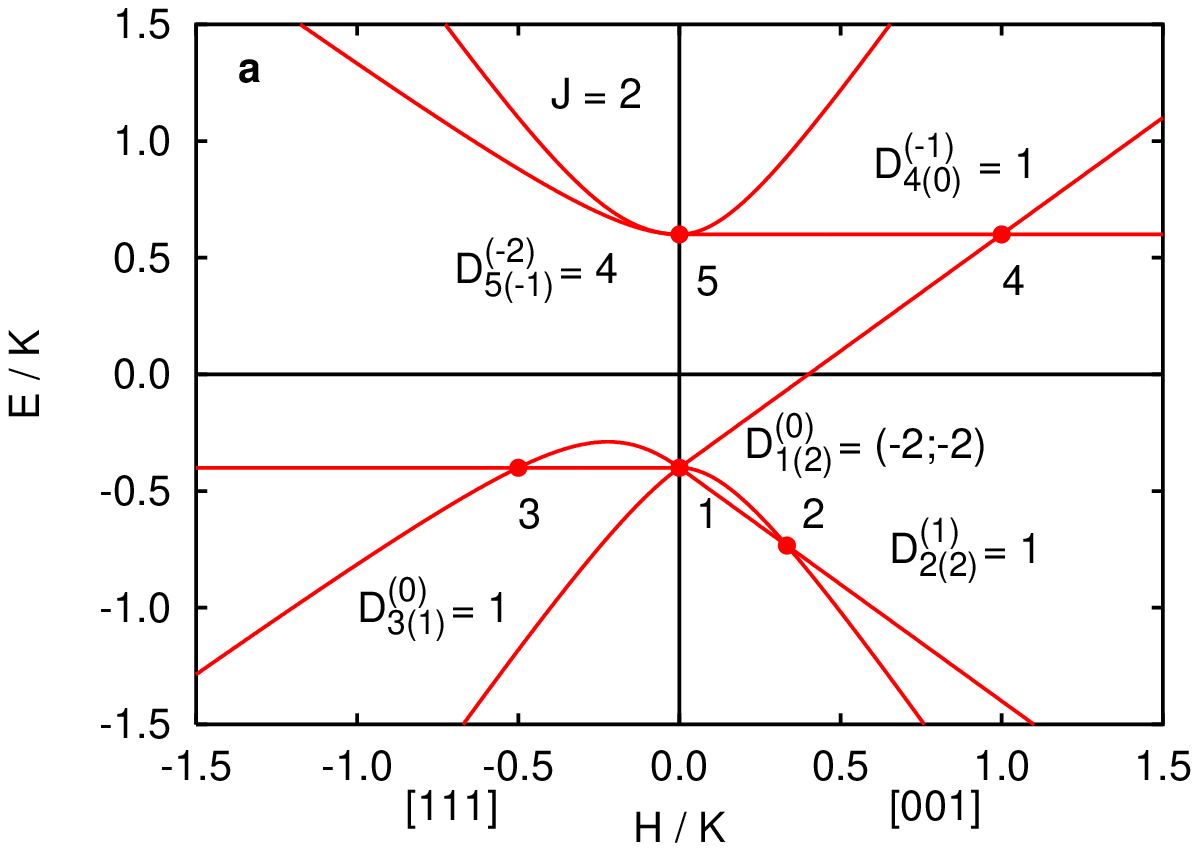} \\
\includegraphics[width=0.8\columnwidth]{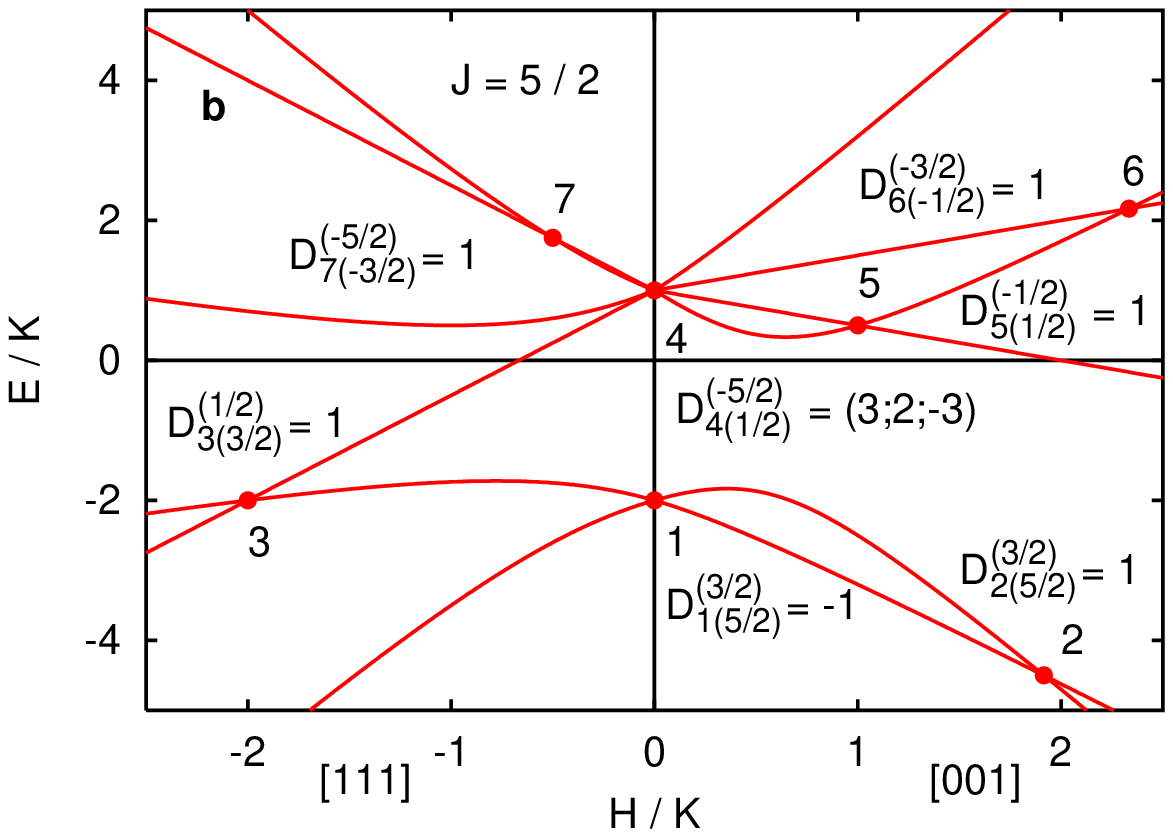}
\end{center}
\caption{Spectrum and diabolical points of a spin $J\!=\!2$
(\textbf{a}) and $J\!=\!5/2$ (\textbf{b}) with cubic anisotropy:
$\mathcal{\hat{H}}_0 \! \equiv\! E_0+
K(\hat{S}_x^4+\hat{S}_y^4+\hat{S}_z^4)/6$. Positive (resp.
negative) values of the field correspond to a field parallel to a
fourfold (resp. threefold) symmetry axis. The diabolical points
are indicates by the solid dots. The corresponding diabolicity
indices are indicated.} \label{fig_cubic}
\end{figure}

The diabolicity index of a diabolical point $i$ for a pair of
successive levels $(\mu, \mu\!-\!1)$ is defined as the sum of the
topological charges up to level $\mu$, i.e.:
$\mathcal{D}_{i(\mu)}^{(\mu-1)} \!\equiv\! \sum_{\mu'\geq \mu}
Q_{i(\mu)} .$
For notation convenience, for a $g$-diabolical point with
$g\!>\!2$, $\mathbf{H}_{i\,(\mu)}^{(\mu+1-g)}$,
we lump the corresponding diabolicity indices into the multiplet
$\mathcal{D}_{i\,(\mu)}^{(\mu-g+1)} \!\equiv\! \left(
\mathcal{D}_{i\,(\mu)}^{(\mu-1)};
\mathcal{D}_{i\,(\mu-1)}^{(\mu-2)};\ldots ;
\mathcal{D}_{i\,(\mu-g+2)}^{(\mu-g+1)} \right)$.
From the latter sum rule for the topological charges, above, we
obtain the following sum rules for the diabolicity indices:
\begin{subequations}
\begin{eqnarray}\label{eq_sum_rule}
\mathcal{D}_{(\mu)}^{(\mu-1)}&\equiv& \sum_i
\mathcal{D}_{i\,(\mu)}^{(\mu-1)}=(J+\mu)\left(J-(\mu-1) \right) , \\
\mathcal{D}&\equiv& \sum_\mu \mathcal{D}_{(\mu)}^{(\mu-1)} =
\frac{2J(J+1)(2J+1)}{3} .
\end{eqnarray}
\end{subequations}
In addition to these sum rules, the set of diabolical points, for
a given Hamiltonian
$\mathcal{\hat{H}}_0$,
must possess all symmetries of
$\mathcal{\hat{H}}_0$;
in particular, the
time-reversal invariance of
$\mathcal{\hat{H}}_0$
implies the
inversion symmetry of the set of diabolical points. One observes
immediately that all these rules are obeyed for the diabolical
points of the biaxial system (Eqs.~(\ref{eq_diabo_biaxial},b)),
where only 2-diabolical points with diabolicity index 1 are found.
Finally, when scaling the Hamiltonian as
$\mathcal{\hat{H}}_0
\rightarrow \lambda\mathcal{\hat{H}}_0$ with $\lambda>0$,
the diabolical points scale as
$\mathbf{H}_{i(\mu)}^{(\mu')}
\rightarrow \lambda\mathbf{H}_{i(\mu)}^{(\mu')}$,
the diabolicity indices remaining unchanged, whereas under
reversing the sign of the Hamiltonian, i.e.
$\mathcal{\hat{H}}_0
\rightarrow -\mathcal{\hat{H}}_0$,
the diabolical points and diabolicity indices change as
$\mathbf{H}_{i(\mu)}^{(\mu')} \rightarrow
\mathbf{H}_{i(-\mu')}^{(-\mu)}$
and
$\mathcal{D}_{i\,(\mu)}^{(\mu-1)} \!\rightarrow
-\mathcal{D}_{i\,(1-\mu)}^{(-\mu)}$.
To illustrate these rules for a case where higher-order diabolical
points occur, I show in Fig.~\ref{fig_cubic} the spectrum and
diabolical points for spins 2 and $5/2$ with cubic anisotropy,
where a rich variety of diabolical points is obtained. The
diabolicity indices characterize the energy dispersion near a
diabolical point: the double-cone (diabolo) shape is obtained only
for a 2-diabolical point of diabolicity index $\pm 1$; otherwise,
a different dispersion law is obtained, as seen in
Fig.\ref{fig_cubic}. The diabolicity index also influences the
nature of the Landau-Zener tunnelling taking place at a diabolical
point. A systematic study of these issues will be given elsewhere.

\begin{figure}[b]
\begin{center}
\includegraphics[width=1.0\columnwidth]{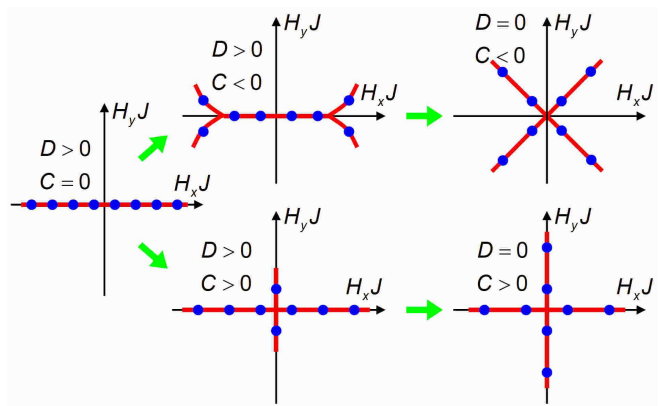}
\end{center}
\caption{Schematic representation of the evolution of diabolical
point distribution (for the two lowest states, $J\!=\!4$), as the
anisotropy progressively changes from biaxial to quadratic.}
\label{fig_bifurcation}
\end{figure}

I then address the above mentioned paradox of the missing
diabolical points for Fe$_8$. As already indicated, it has been
found that that the experimental observations are well explained
quantitatively by adding to the main biaxial Hamiltonian a (very
small) fourth order tetragonal anisotropy term
$\mathcal{\hat{H}}'\equiv C \!\left( \hat{J}_+^4+\hat{J}_-^4
\right)$,
with
$C\!<\!0$
\cite{Wernsdorfer1999}. It has been suggested
\cite{Kececioglu2002} that the additional anisotropy term might
lead to a singular behavior; as we shall see, this explanation is
both correct and incomplete. It is incomplete, because the sum
rule (\ref{eq_sum_rule}) would be violated if the diabolical would
have simply disappeared. Let us qualitatively discuss what happens
as one continuously switches from a biaxial anisotropy
($D\!>\!0$, $C\!=\!0$)
to a tetragonal one
($D\!=\!0$, $C\!<\!0$)
(Fig.~\ref{fig_bifurcation}). The effect of the additional term
(with $C\!<\!0$) is to introduce a new tunnelling path (yellow
arrow in Fig.~\ref{fig_diabo}\textbf{d}); for small values of
$|C|$ and $H_x$, the corresponding amplitude is negligible and the
effect is only to displace the diabolical points along the hard
axis, reducing the distance between the last ones. At a critical
value of $C$, beyond which the amplitude of the yellow path
becomes larger than the one of the red ones, the last 2 diabolical
points collide and a bifurcation takes place. Beyond that point,
the 2 diabolical points symmetrically diverge away from the
$\mathbf{x}$ axis, towards the $\mathbf{x\!+\!y}$ axis (hard axis
for tetragonal anisotropy with $C\!<\!0$); this process then
repeats until all diabolical points have moved to the
$\mathbf{x\!+\!y}$ axis, for $D\!=\!0$. The scenario corresponding
to $C\!>\!0$ is also shown in Fig.~\ref{fig_bifurcation}. For the
case of Fe$_8$ ($J\!=\!10$), this interpretation implies that 3
diabolical points should be located on each branch of the fork
seen for $D>0$ and $C<0$ in Fig.~\ref{fig_bifurcation}. The
experimental check of this prediction would allow to confirm the
present topological theory of diabolical points. Finally, I
propose the following conjecture: \emph{a spin Hamiltonian}
$\mathcal{\hat{H}}_0$ \emph{is completely determined by the set of
its diabolical points and diabolicity indices (together with the
value of its trace)}.

I now come to the last point of this paper, namely the Berry phase
interpretation of the diabolical points found at non-zero values
of $H_z$ for the biaxial system. The difficulty lies in the fact
that the initial ($M$) and final state ($-M'$) of the tunnelling
paths do not generally belong to the same set of coherent states,
which makes the path integral approach of Ref.~\cite{Garg1993}
impracticable. A solution to this difficulty consists in enlarging
the Hilbert space to comprise all possible states of a system of
$2J$ spins $1/2$:
$\mathbf{j}_1,\mathbf{j}_2, \ldots, \mathbf{j}_{2J}$.
The Hamiltonian $\mathcal{\hat{H}}$ operates in this new Hilbert
space by interpreting
$\mathbf{\hat{J}}$
as
$\mathbf{\hat{J}}
\equiv\sum_{i=1}^{2J} \mathbf{\hat{j}}_i$.
We can ensure that the
physics of our problem is thereby unchanged by adding to
$\mathcal{\hat{H}}$
a penalty term
$\mathcal{\hat{H}}'\equiv
-\alpha \left[ \mathbf{\hat{J}}^2 - J(J+1) \right]$
with
$\alpha\rightarrow +\infty$ and considering only the $2J+1$ lowest
levels.

To study the exchange splitting between any pair of states, we
need the following matrix element of the imaginary-time propagator
between two coherent states:
$A\equiv \left\langle
JM\mathbf{n}\right| \mathrm{e}^{-\mathcal{\hat{H}}T} \left|
JM'\mathbf{n}'\right\rangle$ (we set $\hbar =1$).
The coherent states are defined as usual by rotating a state
$\left| JM \right\rangle$ from the $\mathbf{z}$
axis to
$\mathbf{n}\equiv (\sin\theta \cos\varphi, \sin\theta \sin
\varphi, \cos\theta)$,
i.e.,
$\left|JM \mathbf{n} \right\rangle \equiv
\mathrm{e}^{-\mathrm{i}\varphi \mathbf{\hat{J}}_z}
\mathrm{e}^{-\mathrm{i}\theta \mathbf{\hat{J}}_y}
\mathrm{e}^{\mathrm{i}\varphi \mathbf{\hat{J}}z} \left| JM
\right\rangle$.
Clearly, $A$ is unaffected by the penalty term
$\mathcal{\hat{H}}'$ and we can simply omit it for our problem.

One can show that
$\left\langle JM\mathbf{n}\right| \left[\left|
jj\mathbf{n}\right\rangle \otimes \left|
J-j,M-j,\mathbf{n}\right\rangle
\right]=\sqrt{\frac{\left(2J-2j\right)!
\left(J+M\right)!}{\left(2J\right)! \left(J+M-2j\right)!}} \simeq
\left(1- \frac{j(J-M)}{2J}\right)$,
(where the first equality is an exact result, and the second one
an approximation valid for $j(J-M)\ll 2J$), so that
$\left|
JM\mathbf{n}\right\rangle \approx \left| jj\mathbf{n}\right\rangle
\otimes \left| J-j,M-j,\mathbf{n}\right\rangle$
to relative order
$\frac{j(J-M)}{2J}$.
The proof of above equality, to be detailed
elsewhere, uses the fact that states of spin $J$ can be expressed
as completely symmetrized (over all possible permutations)
tensorial products of $2J$ spin-$1/2$ states, and exploits the
group structure of permutations. By a similar argument (together
with the fact
$\mathcal{\hat{H}}$,
depending only on the total spin $\mathbf{\hat{J}}$, commutes with
the permutation operator), one can also prove the following exact
result:
$A\!=\! \sqrt{\!\frac{\left(2J\right)!
\left(J-M-2j\right)!}{\left(2J-2j\right)! \left(J-M\right)!}}
\left[ \left\langle jj,\!-\mathbf{n} \right|\!\otimes\!
\left\langle J\!-\!j,M\!\!+\!j,\!\mathbf{n}\right| \right]
\mathrm{e}^{-\hat{H}T} \!\left| JM^\prime \mathbf{n}^{\prime}
\right\rangle$. Combining those results, we obtain:
$A \! \propto \!\left[\left\langle jj,\!-\mathbf{n}\right|\!
\otimes\! \left\langle
\tilde{J}\tilde{M}\mathbf{n}\right|\right]\!\mathrm{e}^{-\hat{H}T}\!
\left[\left| jj\mathbf{n}'\right\rangle\!\otimes\!
\left|\tilde{J}\tilde{M}\mathbf{n}'\right\rangle\right]$,
with
$j\!\equiv\!(M'\!-\!M)/2$
(without restriction, we assume
$M'\!\geq\! M$), $\tilde{J}\!\equiv\! J\!-\!j$, and
$\tilde{M}\!\equiv\! (M\!+\!M')/2$.
Now, for large values of $J$ and small values of the applied field
along the hard axis,
$\mathbf{n}$ and $\mathbf{n}'$
remain very
close to
$\mathbf{z}$ and $-\mathbf{z}$,
respectively, so that we can write
$\left|jj,-\mathbf{n}\right\rangle \approx
\mathrm{e}^{\mathrm{i}\alpha} \left|j,-j\right\rangle$
and
$\left|jj,\mathbf{n}'\right\rangle \approx
\mathrm{e}^{\mathrm{i}\alpha'} \left|j,-j\right\rangle$.
This finally gives
$A\propto \int\mathcal{D}\mathbf{u}(\tau)
\mathrm{e}^{-\mathcal{S}[\mathbf{u}(\tau)]}$,
where the path integral is for coherent states
$\left|
\tilde{J}\tilde{M}\mathbf{u}\right\rangle$
with
$\mathbf{u}(0)\equiv \mathbf{n}$
and
$\mathbf{u}(T)\equiv
\mathbf{n}'$,
and where the action is given, as usual, by
$\mathcal{S}[\mathbf{u}(\tau)]\equiv
\mathcal{S}_{\mathrm{WZ}}[\mathbf{u}(\tau)] +
\mathcal{S}_H[\mathbf{u}(\tau)]$.
The first term is the Wess-Zumino (or Berry phase) action,
$\mathcal{S}_{\mathrm{WZ}}[\mathbf{u}(\tau)] \equiv
\mathrm{i}\tilde{M} \int \left(1-\cos\theta_\mathbf{u} \right)
\mathrm{d}\varphi_\mathbf{u}$,
responsible for the quantum interferences \cite{Loss1992}; the
second term is the dynamical action,
$\mathcal{S}_H[\mathbf{u}(\tau)]\equiv \int_0^T
\mathrm{d}\tau\, E(\mathbf{u}(\tau))$,
where the energy is
$E(\mathbf{u}) \equiv \left[\left\langle j,-j\right|\otimes
\left\langle \tilde{J}\tilde{M}\mathbf{u}\right| \right]
\mathcal{\hat{H}} \left[\left| j,-j\right\rangle\otimes \left|
\tilde{J}\tilde{M}\mathbf{u}\right\rangle \right]$.
In short, we have mapped our original problem onto that of the
tunnelling between the states
$\left|
\tilde{J}\tilde{M}\mathbf{n}\right\rangle$
and
$\left|
\tilde{J}\tilde{M}\mathbf{n}'\right\rangle$
of a fictitious spin
$\tilde{J}$
with biaxial anisotropy, which can be treated by the
instanton method as in Ref.~\cite{Garg1993}. Noting that this
fictitious spin is subject to the effective field
$H_z^{\mathrm{eff}}\equiv H_z -2jK$,
along the easy axis, we immediately generate the \emph{complete}
set of diabolical points (\ref{eq_diabo_biaxial},b), with $H_z^0
\equiv K$; this agrees well with the exact result
$H_z^0 \equiv \sqrt{K^2-D^2}$
for
$D\ll K$,
which is actually the case for Fe$_8$. By mapping the original
tunnelling problem onto that of a fictitious spin in an effective
$H_z$ field, we obtain a simple interpretation of all the
diabolical points in terms of destructive interferences due to the
Berry phase for the effective spin $\tilde{J}$. The striking
parity alternation discovered by Wernsdorfer and Sessoli
\cite{Wernsdorfer1999} (red \emph{vs.} blue points in
Fig.~\ref{fig_diabo}c), is thus simply interpreted as due to
$\tilde{J}$ being alternately integer and half-integer.

Note, that in principle, our approach is supposed to be valid only
in the limit of large $J$ and for small $H_x$ and $H_z$; it is
thus a surprise to see that it essentially yields exact results,
even for small $J$ and/or for large $H_x$ and $H_z$. This puzzle
has already been noticed \cite{Villain2000, Kececioglu2001} and is
not fully understood.

\end{document}